\documentclass[useAMS,usegraphicx,usenatbib]{mn2e}
\newcommand{\kms}{~km~s$^{-1}$}

\newcommand{\chisq}{$\chi^{2}$}
\newcommand{\vsini}{$v\sin i$}

\newcommand{\ha}{{\mbox{H$\rm \alpha$}}}

\title[Dynamo Processes in the T~Tauri star V410~Tau]{Dynamo Processes in the T~Tauri star V410~Tau\thanks{Based on observations obtained at the T\'elescope Bernard Lyot 
  (TBL), operated by CNRS/INSU.}}
\author[Mairead B. Skelly]{M.B. Skelly$^{1}$\thanks{E-mail:
mskelly@ast.obs-mip.fr}, J.-F. Donati$^{1}$, J. Bouvier$^{2}$, K.N. Grankin$^{3}$, Y.C.~Unruh$^{4}$, 
\newauthor
S.A.~Artemenko$^{3}$, P.Petrov$^{3}$  \\
$^{1}$ Observatoire Midi-Pyrenees, LATT, 14 avenue Edouard Belin, Toulouse, 31400, France \\
$^{2}$  Laboratoire d'Astrophysique de Grenoble, Observatoire de Grenoble, BP 53, 38041, Grenoble, France \\
$^{3}$  Crimean Astrophysical Observatory, Nauchny, Crimea 334413, Ukraine \\
$^{4}$ Astrophysics Group, Imperial College of Science, Technology and Medicine, London SW7 2AZ}

\begin{document}

\date{11th November 2009}

\pagerange{\pageref{firstpage}--\pageref{lastpage}} \pubyear{2009}

\maketitle

\label{firstpage}

\begin{abstract}

\noindent We present new brightness and magnetic images of the weak-line T~Tauri star V410~Tau, made using data from the NARVAL spectropolarimeter at T\'{e}lescope Bernard Lyot (TBL). The brightness image shows a large polar spot and significant spot coverage at lower latitudes. The magnetic maps show a field that is predominantly dipolar and non-axisymmetric with a strong azimuthal component. The field is 50\% poloidal and 50\% toroidal, and there is very little differential rotation apparent from the magnetic images. 

A photometric monitoring campaign on this star has previously revealed V--band variability of up to 0.6 magnitudes but in 2009 the lightcurve is much flatter. The Doppler image presented here is consistent with this low variability. Calculating the flux predicted by the mapped spot distribution gives an peak-to-peak variability of 0.04 magnitudes. The reduction in the amplitude of the lightcurve, compared with previous observations, appears to be related to a change in the distribution of the spots, rather than the number or area.  

This paper is the first from a Zeeman-Doppler imaging campaign being carried out on V410~Tau between 2009--2012 at TBL. During this time it is expected that the lightcurve will return to a high amplitude state, allowing us to ascertain whether the photometric changes are accompanied by a change in the magnetic field topology. 

\end{abstract}


\section{Introduction}

T~Tauri stars (TTS) are pre-main sequence late-type stars. As they contract towards the main sequence the increasing density in their interiors will lead to the development of a radiative core, if the star is sufficiently massive. A point of uncertainty concerns the evolution of their dynamos. In fully convective stars, or stars with small radiative cores, we expect the dynamo to look fundamentally different to that of the Sun. We presume that the solar dynamo is concentrated in the boundary between the convective and radiative layers, known as the tachocline. As the radiative core in the young star grows the dynamo should change from being distributed throughout the convective zone to being confined to the tachocline. Observations of young stars with a range of ages, masses and rotation rates will help to illuminate this process. 

V410~Tau is one of the most well-observed TTS. It is a weak-line T~Tauri star (wTTS) which has already dissipated its disk, hence it does not display signatures of accretion characteristic of classical TTS (cTTS). It has a \vsini\ of $74\pm3$\kms\ and a period of 1.872~days \citep{stelzer03}, and is therefore an example of the young, rapidly rotating stars which are characterised by strong, large scale magnetic fields. Other, more evolved, examples include the 10~Myr old TWA~6 \citep{skelly1}, Speedy~Mic \citep[20 Myr,][]{barnes01,barnes05c, duns06} and AB~Dor \citep[50~Myr,][]{cc02b,donati03b,jeffers07}. 

The strong fields in these stars have either been directly measured using Zeeman-Doppler imaging \citep[ZDI,][]{semel89,jf97,tausco06}, or their presence indicated indirectly e.g. by the large spot coverage on the surface. In the case of V410~Tau this field has revealed itself in the presence of cool spots covering a large fraction of the photosphere. In fact, V410~Tau has exhibited one of the highest optical variabilities observed on a young star, with $\rm \Delta V$ reaching 0.6 magnitudes \citep{stelzer03,grankin08}.

V410~Tau is the target of an ongoing photometric monitoring campaign which began in 1981 \citep{vrba88, herbst89,petrov94, grankin08}. For most of this time the lightcurve has been smooth, repeating, with a clearly defined period and an amplitude between 0.2 and 0.6 magnitudes. 

Two significant decreases in the amplitude of the lightcurve, where $\Delta \rm V$ fell below 0.2, have been observed during this campaign. The first of these was between 1981--1983 and the second began in 2005, reaching a minimum in 2007. Additional observations of V410~Tau dating from 1905--1987 are available on photographic plates in the archives of the Sternberg Astronomy Institute. \cite{sokoloff08} have digitised this data and used it to investigate the variability over a longer period of time. This revealed a third decrease in lightcurve amplitude between 1963--1970, this time in the $B$--band. 

These changes in the lightcurves suggest a rearrangement of the cool spots.  
Given that starspots are a magnetic phenomenon such a change is likely to be caused by a change in the magnetic topology. The fact that this change has been observed three times with intervals of approximately 20~years may be evidence of a stellar cycle, similar to the 22-year solar cycle. The existence of magnetic cycles in active stars has been extensively investigated through changes in their starspot coverage \cite[see, e.g., ][and references therein]{berd05, strass09}. Possible short-period cycles (of between 5--7 years) have been found on V410~Tau by \cite{stelzer03} and \cite{olah09}, using photometric data spanning several decades.

Doppler imaging \citep{vogt87,strass02} can provide more detail on the changes in spot distribution. Maps of the spot distribution on V410~Tau have been made by \cite{v41094b, hatzes95b} and \cite{v41096}. The images have shown decentred polar spots and low latitude spots. Often the lower latitudes features are found to be confined within particular longitude ranges \citep[e.g.][]{v41096}. 

The flattening of the lightcurve suggests that a change in the distribution or number of photospheric spots has occurred, i.e., either there are fewer spots than previously or there is a more even longitudinal distribution of spots. \cite{grankin08} found a stronger correlation between the non-uniformity of the spot distribution and the amplitude of the lightcurve than the total spot area. This suggests that the reduction in lightcurve amplitude is related to a rearrangement in the spot distribution, rather than a change in the number or area of the spots. 

To date, no magnetic images of V410~Tau have been published. This has motivated a long--term campaign, being carried out at T\'{e}lescope Bernard Lyot (TBL), which aims to produce several spot and magnetic maps of V410~Tau between 2009--2012. We expect that during this time the lightcurve will return to the state observed between 1988--2004. 
In this paper we present the first set of images. Sec.~2 discusses the evolutionary status of V410~Tau. Sec.~3 describes the observations and data reduction; in Sec.~4 we describe the image production and present the brightness and magnetic images, and a new measurement of the differential rotation. Sec.~5 is a discussion of the implication of the results for the dynamo theory in pre-main sequence stars. 

\section{Physical parameters of V410~Tau}\label{phys}

The evolutionary state of V410~Tau is somewhat uncertain. The \vsini\ and period quoted above indicate a minimum stellar radius of $2.8~\rm R_{\odot}$. Widely varying effective temperatures have been reported in the literature, including 4700~K \citep[e.g.,][]{bertout07} and 4060~K \citep[e.g.,][]{sestito08}. In this work we adopt a temperature of $4500\pm100~\rm K$, in better agreement with the B - V value of 1.16 from \cite{grankin08} and the temperature calibrations of \cite{bessel98}. 

With this temperature and minimum radius we can estimate the minimum luminosity to be $3.4 \rm~L_{\odot}$. \cite{wichmann98} and \cite{bertout99} give a range of distances of 116--157~pc, based on $Hipparcos$ \citep{hipparcos} measurements.  Assuming the lowest observed V magnitude (10.52) in \cite{grankin99} is the unspotted V magnitude, and a distance of 157~pc gives a luminosity of $2.4 \rm~L_{\odot}$ (if $\rm A_{v}=0$), well below the minimum luminosity calculated above.  

The discrepancy suggests that even when we observe the star at its brightest, it still has significant spot coverage. In other words, the true ``unspotted'' magnitude is brighter than 10.52. This possibility was explored in \cite{grankin99}, who calculated an absolute unspotted V mag of 10.236, 0.28 mag brighter than their minimum observed magnitude. This would mean that the star has an even distribution of small spots across its surface at all times that are not detected by Doppler imaging, as they are smaller than the resolution limit.

Using an inclination of $70\pm10^{\circ}$ (obtained in the imaging process, see Sec.~\ref{bright}) gives a radius of $\rm3.0~R_{\odot}$, and with a temperature of 4500~K indicates \citep[using the models of][]{siess00} that V410~Tau has a mass of $1.4\pm0.2~M_{\odot}$ and an age of $1.2\pm0.3$~Myr. This is much younger than previously indicated, e.g. in  \cite{bertout07}, where an age of 6~Myr was given for V410~Tau. The parameters given in \cite{bertout07} are consistent with a radius of $1.8 \rm~R_{\odot}$, which is incompatible with the \vsini\ and period. 

These parameters indicate \citep[again using the models of ][]{siess00} that the radiative core has a radius of $0.1 \rm R_{*}$. The uncertainties in the temperature and inclination allow the radiative core to have a radius between $0.0-0.28 \rm R_{*}$, meaning that the star could be fully convective. 
 
Some interesting questions arise when we consider the possibility of a stellar cycle, particularly as we expect that a star which is fully convective or has a small radiative core has close to `solid body' rotation \citep{barnes05}. 

Differential rotation is crucial to the solar magnetic cycle, as it allows the development of toroidal field from poloidal field by shearing the field lines \citep{parker55}. This may not rule out the possibility of cyclical dynamos existing without differential rotation. For example, \cite{rudiger03} modelled $\rm \alpha^{2}$ dynamos and found that oscillatory solutions appear in turbulent spherical shells in exceptional cases. However, they conclude that cyclical activity is almost always indicative of differential rotation.  

Previous attempts to detect differential rotation on V410~Tau have found it to be very small. \cite{v41096} obtained a value for the differential rotation parameter $\Delta\Omega$ of $0.003$, around 1/20 of the solar value. If the star does indeed have a cycle it appears this must arise with very little differential rotation. 

\section[]{Observations and Data Analysis}\label{obs}

Spectroscopic observations of V410~Tau were made at the 2-m T\'{e}lescope Bernard Lyot (TBL) on Pic du Midi in the French Pyr\'{e}n\'{e}es between 02--17 January 2009. During this time, 40 spectra were taken using the spectropolarimeter NARVAL (the twin of ESPADONS on the Canada France Hawaii Telescope), with four sub-exposures taken in different polarimeter configurations each time. This allows spurious polarisation signals to be removed to first order \citep{jf97}. The integration times (for the individual sub-exposures) were 600s, except for the observations taken on 10 January 2009 which had integration times of 800s. 
 
The full wavelength range of the spectra was $370-1000~\rm nm$, taken with a spectral resolution of 60000 and a velocity pixel step of 2.6\kms, reduced to approximately 1.8\kms\ as the spectrograph slit was tilted with respect to the CCD pixels. The full observing log is given in Tab.~\ref{obs1}. In the table the rotational phase of V410~Tau at each observation is given with respect to the ephemeris,
\begin{equation}
\rm JD(Hel.) = 2454832.58033 + 1.87197R, 
\end{equation}

\noindent where R is an integer number of  rotations. This uses the same period and zero phase as \cite{stelzer03}. The resulting phase coverage is good, covering around 75\% of the surface and there is close phase correspondence between several of the observations providing a good opportunity to measure differential rotation (see Sec.~\ref{diffrot}). 
 
\begin{table}
\centering
\caption{Spectra taken at NARVAL/TBL. The dates are all in January 2009. The observing times shown indicate the beginning of the integration. Column 4 shows the maximum signal-to-noise ratio per CCD pixel.  Column 5 is the rms noise level of the polarised LSD profiles.  }
\begin{tabular*}{0.48\textwidth}{@{\extracolsep{\fill}}lllrcr}
\hline 
Date & HJD & UT & S/N & $\rm \sigma_{LSD}$ & Rotation \\
 & 2454800+ & (h:m:s) & & [$10^{-4}$] & cycle \\
\hline
02 & 34.32160 & 19:38:17 & 166 & 2.12 & 0.9302 \\
02 & 34.35298 & 20:23:28 & 164 & 2.10 & 0.9470 \\
02 & 34.38435 & 21:08:39 & 166 &  2.08 & 0.9637 \\
02 & 34.41572 & 21:53:50 & 166 &  2.07 & 0.9804 \\
02 & 34.44857 & 22:41:08 & 158 &  2.16 & 0.9980 \\
03 & 35.30280 & 19:11:19 & 158 &  2.24 & 1.4544 \\
03 & 35.33417 & 19:56:30 & 159 &  2.13 & 1.4712 \\
03 & 35.36601 & 20:42:21 & 159 &  2.17 & 1.4882 \\
03 & 35.39786 & 21:28:13 & 151 &  2.30 & 1.5052 \\
03 & 35.42926 & 22:13:25 & 146 &  2.31 & 1.5220 \\
04 & 36.28992 & 18:52:52 & 148 & 2.45 & 1.9818 \\
04 & 36.32130 & 19:38:04 & 155 &  2.29 & 1.9985 \\
04 & 36.35268 & 20:23:16 & 157 &  2.17 & 2.0153 \\
04 & 36.38407 & 21:08:28 & 156 &  2.24 & 2.0320 \\
04 & 36.41546 & 21:53:40 & 157 &  2.30 & 2.0488 \\
10 & 42.36738 & 20:45:05 & 157 &  2.58  & 5.2286 \\
10 & 42.40803 & 21:43:38 & 157 & 2.30 & 5.2502 \\
10 & 42.44867 & 22:42:09 & 164 & 2.31 & 5.2720 \\
11 & 43.29158 & 18:56:02 & 135 & 2.79 & 5.7224 \\
11 & 43.32298 & 19:41:16 & 141 & 2.63 & 5.7391 \\
11 & 43.35436 & 20:26:27 & 129 & 2.95 & 5.7559 \\
11 & 43.38575 & 21:11:39 & 115 &  3.43 & 5.7727 \\
12 & 44.32276 & 19:41:03 & 115 & 3.54 & 6.2733 \\
12 & 44.35416 & 20:26:17 & 113 & 3.49 & 6.2900 \\
12 & 44.38555 & 21:11:29 & 115 &  3.74 & 6.3068 \\
12 & 44.41998 & 22:01:04 & 130 & 2.84 & 6.3251 \\
14 & 46.30265 & 19:12:20 & 143 & 2.65 & 7.3310 \\
14 & 46.33402 & 19:57:30 & 145 & 2.56 & 7.3478 \\
14 & 46.36540 & 20:42:42 & 145 & 2.48 & 7.3645 \\
14 & 46.39678 & 21:27:54 & 145 & 2.49 & 7.3812 \\
16 & 48.26894 & 18:24:01 & 136 &  2.72 & 8.3815 \\
16 & 48.30032 & 19:09:13 & 149 &  2.43 & 8.3983 \\
16 & 48.33172 & 19:54:26 & 144 & 2.50 & 8.4150 \\
16 & 48.36311 & 20:39:38 & 131 & 2.83 & 8.4318 \\
16 & 48.43488 & 22:23:00 & 112 & 3.37 & 8.4701 \\
17 & 49.26621 & 18:20:13 & 126 & 3.01 & 8.9143 \\
17 & 49.29759 & 19:05:25 & 102 & 4.03 & 8.9310 \\
17 & 49.32898 & 19:50:37 & 120 & 3.16 & 8.9478 \\
17 & 49.36037 & 20:35:49 & 139 & 2.72 & 8.9646 \\
17 & 49.39175 & 21:21:01 & 136 & 2.71 & 8.9813 \\
\hline
\end{tabular*}
\label{obs1}
\end{table}

\begin{figure}
\includegraphics[width=0.45\textwidth]{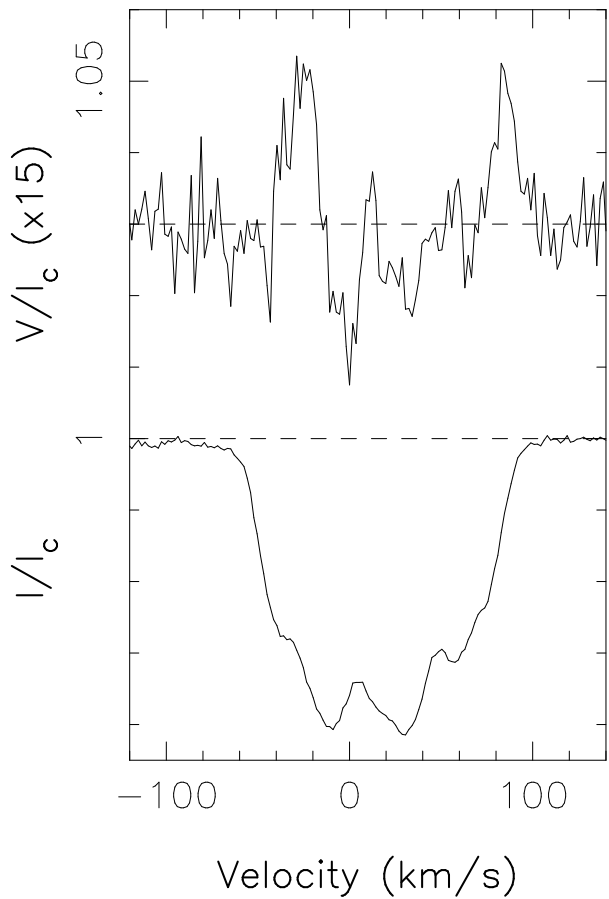}
\caption{Example Stokes I (bottom) and V (top) profiles at phase 7.331. The V profile has been multiplied by fifteen and shifted upward by 1.03.}
\label{exprof}
\end{figure}

The data was reduced using L{\sc ibre}  ES{\sc p}RIT, a dedicated pipeline for the reduction of spectropolarimetric data taken at TBL and CFHT. This package automatically extracts the unpolarised and polarised spectra, as described in \cite{jf97}. To improve the effective signal-to-noise ratio of the data least--squares deconvolution \citep[LSD,][]{jf97} was applied to each spectrum, i.e., a line list was deconvolved from each spectrum, revealing in each case the underlying rotationally broadened profile. Mean intensity (Stokes I) and circularly polarised (Stokes V) profiles were produced in this way. The line list was created using an {\sc atlas}9 model atmosphere \citep{kurucz93}, using a spectral type of K5, and contained just under 9000 absorption lines in total. The Stokes I profiles show distortions characteristic of photospheric spots and Zeeman signatures were detected in most spectra, with a typical peak-to-peak amplitude of 0.2\%, i.e. 10 times the noise level. Example Stokes I and V profiles are shown in Fig.~\ref{exprof}.

\begin{figure}
\centering
\includegraphics[width=0.33\textwidth, angle=270]{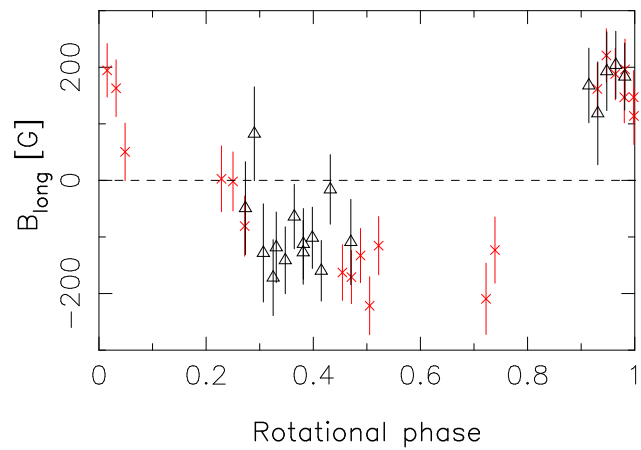}
\caption{Line-of-sight magnetic field calculated using Eq. \ref{eq:b}. The black triangles represent rotations $0-5$ (as listed Tab. \ref{obs1}) while the red crosses represent rotations $6-8$. }
\label{blong}
\end{figure}

We used the I and V profiles to directly calculate the magnetic field along the line-of-sight. This field is calculated using 

\begin{equation}
{\rm B}_{l}(G) = -2.14 \times 10^{11} \frac {\int{\rm v V(\rm v)\rm d\it \rm{v}}}{\lambda_{0}\rm gc \int{[I_{c}-I(v)]{\rm d}v}}
\label{eq:b}
\end{equation}

\noindent from \cite{jf97} where v is the velocity in the star's rest frame, $\lambda_{0}$ the central wavelength of the LSD profile in nanometres, c is the velocity of light in vacuum, $\rm g_{\rm eff}$  is the value of the mean Land\'{e} factor of the LSD line and $\rm I_{c}$ is the continuum level. This is plotted in Fig.~\ref{blong}. Near phase 0.0 the field peaks at a value of around 200G. The value falls to a plateau between phases $0.25 - 0.5$ where the field is about 150~G in the other direction. Although the individual error bars are large the plot reveals the rotational modulation of the longitudinal field, as the pattern of variability just described is repeated throughout the eight rotations that were observed. 
  
Photometric observations of V410 Tau were obtained with two different photoelectric telescopes at different locations in Crimea, Ukraine: the 0.6-m T-60 on Mt. Koshka equipped with UBV(R)c filters for the Johnson-Cousins system, and the 1.25-m AZT-11 in Nauchny equipped with the similar filters. Altogether, 41 UBV(R)c data points were collected for this star between 03 August 2008 and 26 January 2009. All measurements were made differentially with respect to a nearby comparison star and a check star in the the following sequence: CK-S-C-V-C-V-C-V-C-S-CK, where CK is the check star, C the comparison star, S the sky 
background, usually between the comparison and the variable star, and and V is the variable itself. Integration time was usually set to 10s for all filters. The data reduction was based on nightly extinction coefficients obtained from a set of standard stars. The rms error of a single measurement in the instrumental system is about 0.005~mag in BVR and 0.01~mag in U. Only the V-band lightcurve is used in this work.

\section{Doppler and Magnetic Images}
 
\subsection{Zeeman-Doppler imaging process}
 
Doppler imaging produces maps of cool photospheric spots by tracking the distortions that the spots cause to rotationally broadened spectral lines as the star rotates. Zeeman Doppler imaging extends this technique to mapping magnetic fields by applying a similar process to the circular polarisation profiles. 

Details of the Zeeman Doppler imaging process have been given in several papers including \cite{jf97} and \cite{tausco06}. The code finds the best-fitting images using \chisq\ minimisation using the I and V profiles as constraints. This cannot provide a unique solution, as the spatial resolution of the image is always limited to some extent.  For this reason, entropy maximisation is used as an additional constraint, i.e. the information content of the image is minimised.

The ZDI process usually begins with a spotless star and a null magnetic field. At each step in the iteration the I and V profiles corresponding to the current image are calculated. The local line profiles are described using Gaussian profiles, which has been found to be adequate in the case of rapidly rotating stars where the shape of the local line profile is less important. The central wavelength of the mean line is 620~nm, the Land\'{e} g factor of 1.2 and a FWHM of 9.7~\kms.  A linear limb-darkening law with a coefficient of 0.75 was used. The code explores the parameter space, finding stellar configuration which gives $\chi^{2}_{r}=1.0$ while simultaneously maximising the image entropy. 

\subsection{Brightness Image}\label{bright}

The free parameters in the production of the brightness image are the spot-filling factors of each pixel of the image \citep{cc92}, i.e., the fraction of each pixel which contains spot.

V410~Tau is a well-studied star, therefore the stellar parameters, including the \vsini, period, radial velocity, inclination and differential rotation, have been measured in previous work \citep[e.g.,][]{hatzes95, v41096, stelzer03}. We can confirm these values using \chisq\ minimisation before producing the final images. To fit the stellar parameters 1000 image pixels were used, while the final images were produced using 10000 pixels. The final reconstruction was carried out using a radial velocity of 16.8\kms, projected rotational velocity \vsini\ = 75\kms\ and inclination $i = 70^{\circ}$ (consistent with the values in the introduction). The fitting of the differential rotation and period is described in Sec.~\ref{diffrot}.

\begin{figure*}
\includegraphics[width=0.75\textwidth]{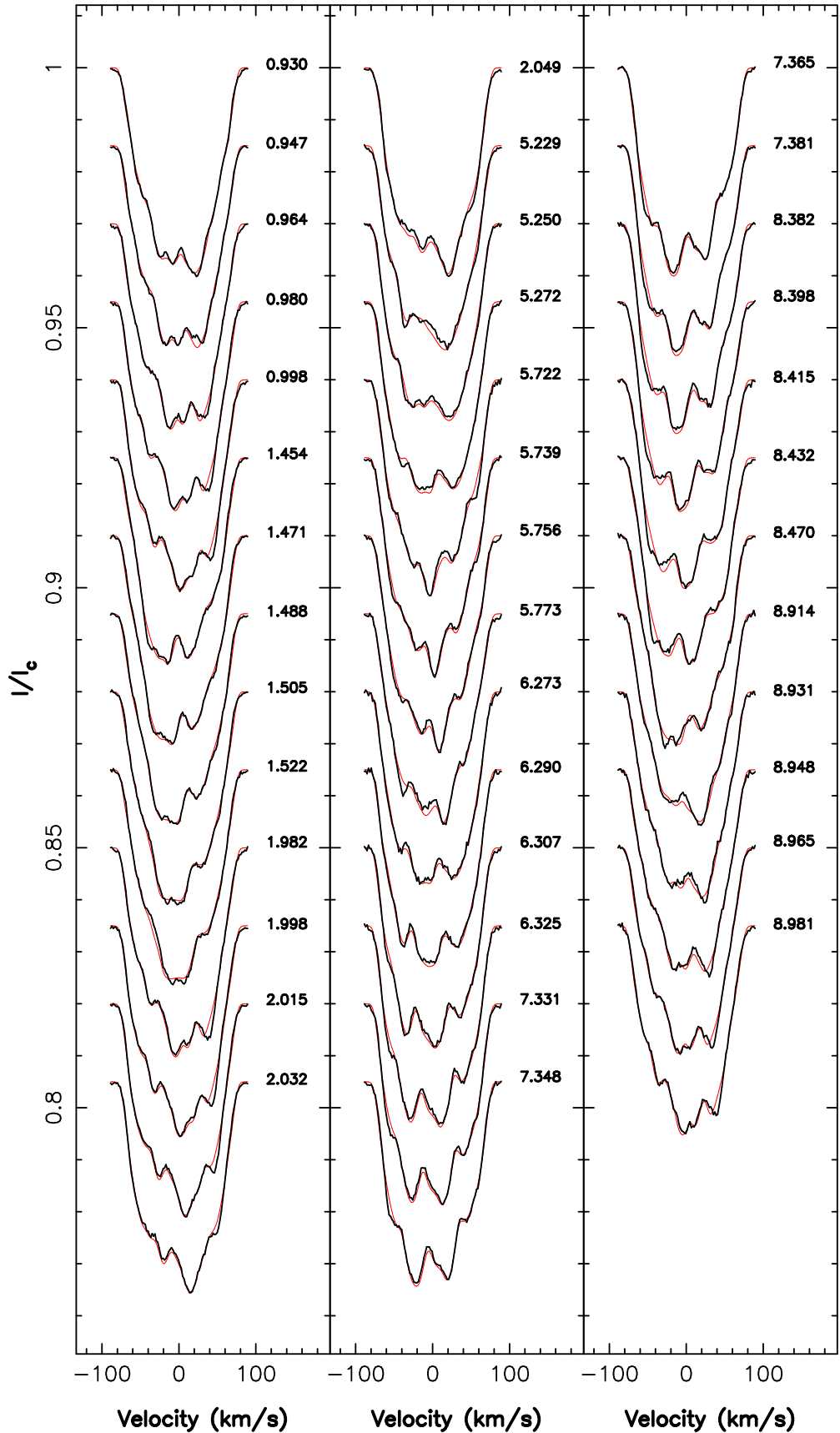}
\caption{The I profiles produced using LSD are plotted in black and the fits are plotted with a thin red line. The number of rotational profiles completed since the ephemeris is given on the right. }
\label{profsi}
\end{figure*}

\begin{figure}
\includegraphics[width=0.5\textwidth,clip=true,trim=0 420 0 0]{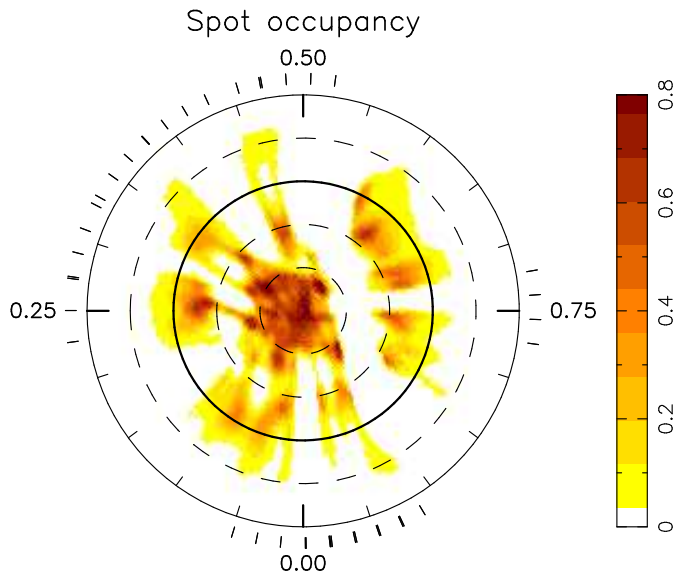}
\caption{Doppler image of V410~Tau shown as a flattened polar projection. The colour scale represents the spot--filling factor. The circles represent lines of latitude separated by $30^{\circ}$, the largest circle is at latitude $-60^{\circ}$ and the bold line is the equator. The  numbers around the outside are rotational phases and the outer tickmarks show phases at which observations were made. } 
\label{dopim}
\end{figure}

Fig.~\ref{profsi} shows the measured intensity profiles and their fits, while the Doppler image 
is shown in Fig.~\ref{dopim}. It reveals a polar spot and groups of spots at lower latitudes. The total spot filling factor on the visible regions of the star is 8\%. The lower latitude spots are concentrated in the latitude band between $0-30^{\circ}$ and are fairly evenly spread longitudinally. The relationship between the spot distribution and the optical variability will be discussed in Sec. \ref{pho}.

\subsection{Magnetic images}

The poloidal and toroidal components of the magnetic field are described using a spherical harmonic expansion with each component of the field (the radial, non-radial poloidal and toroidal field components respectively) expressed in spherical coordinates. The equations describing the field are set out in \cite{tausco06}. The advantage of using a spherical harmonic decomposition, rather than allowing each pixel to vary freely, is that the spherical harmonic description gives rise to more physically realistic solutions. For example, the code will straightforwardly return a dipole, which is unlikely when the field values in individual pixels are left as independent variables. 

\begin{figure*}
\includegraphics[width=0.75\textwidth]{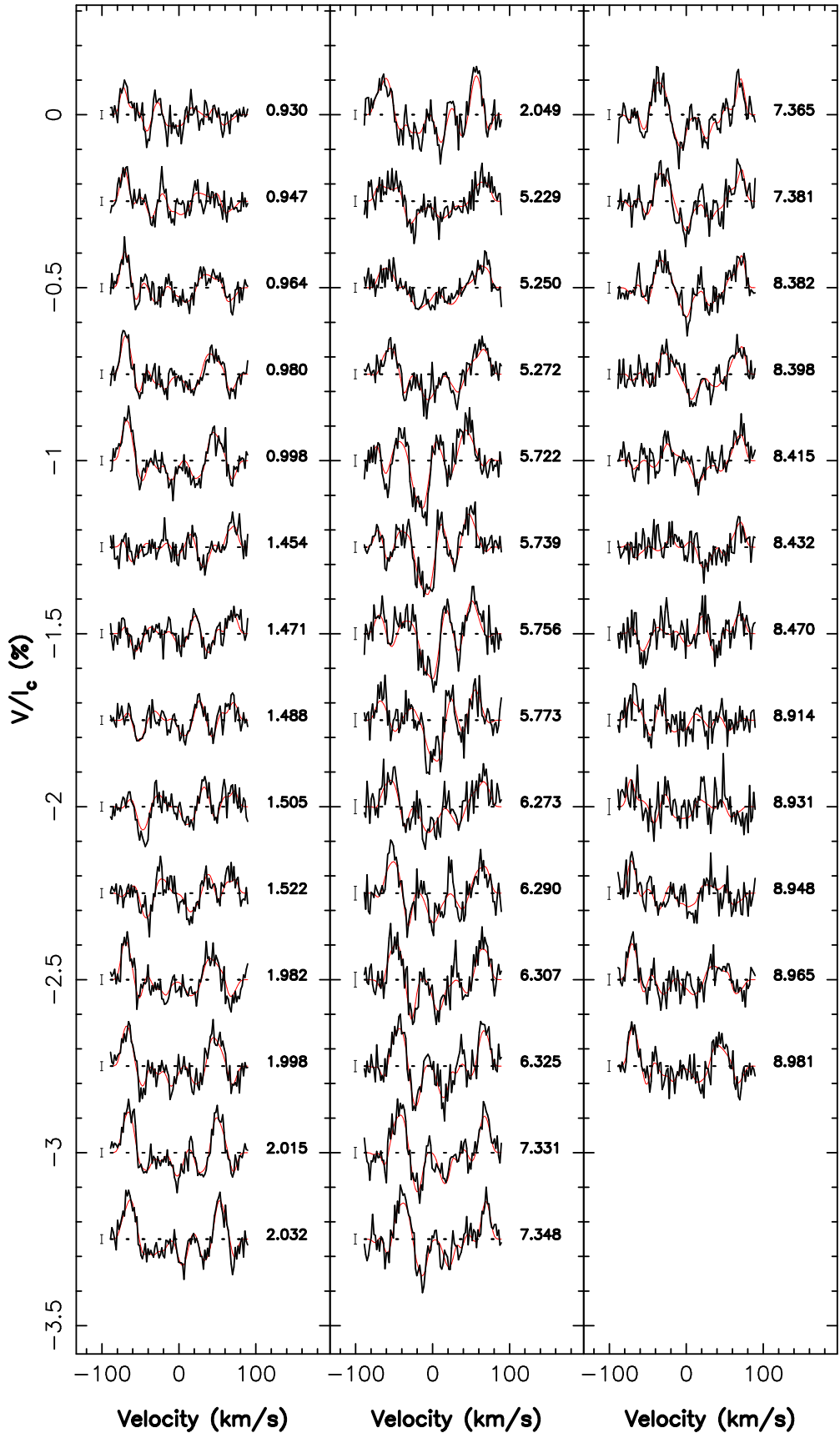}
\caption{As Fig.~\ref{profsi} for V profiles. The number of rotational profiles completed since the ephemeris is given on the right. The 1-sigma error bars are on the left--hand side of each profile.}
\label{profsv}
\end{figure*}

As the spatial resolution depends on the ratio of the width of the absorption lines to the velocity resolution in the spectra, the high \vsini\ of V410~Tau translates to a high resolution on the stellar surface. This allows the field to be fitted up to $l = 15$. Fig.~\ref{profsv} shows the resultant V profiles and the fits. 

\begin{figure}
\includegraphics[width=0.45\textwidth]{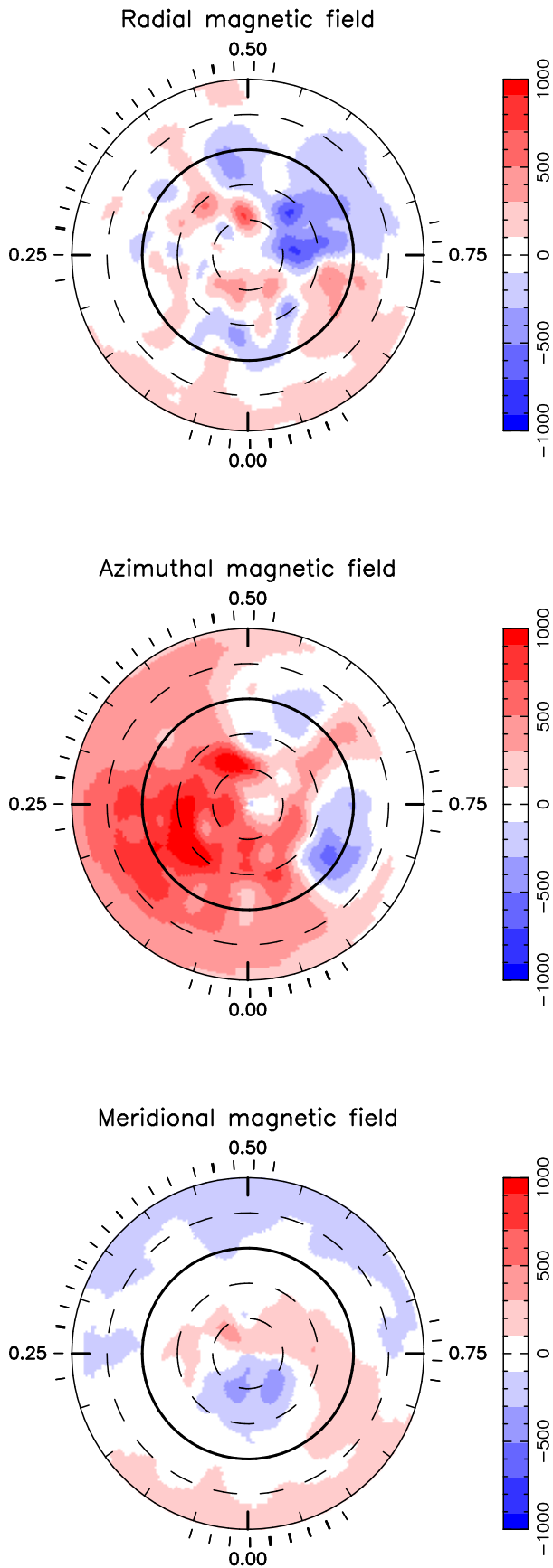}
\caption{The magnetic images of V410~Tau, displayed using the same projection as Fig.~\ref{dopim}. The colour scale represents the magnetic field in units of Gauss. } 
\label{magim}
\end{figure}

The radial, azimuthal and meridional components of the field are shown in Fig.~\ref{magim}. These images show a strong, apparently complicated magnetic field, with individual regions having strengths of up to 1~kG.  The average field integrated over the surface of the star is 490~G. The radial component is complex with field covering a large proportion of the surface, consisting of many small regions of magnetic field and several switches in polarity. The azimuthal field covers almost the entire surface of the star, with field strengths between $300 - 1000~\rm G$ covering more than half the visible surface. There is a monolithic region of positive polarity completely covering one hemisphere, with two smaller regions of negative polarity in the opposite hemisphere. The meridional field is weaker than the radial and azimuthal components and shows positive and negative regions near the pole.

We can quantify the complexity of the field by calculating the fractional strengths of different components. The field is evenly split between the poloidal (49.2\%) and toroidal (50.8\%) components. The poloidal is dominated by the dipolar component, accounting for 60\% of the total field. The poloidal field is nearly completely non-axisymmetric, with 5.3\% of the poloidal field being axisymmetric.
The implications of the magnetic topology in relation to the fields in young stars in general will be discussed in Sec. \ref{dyn}.

\subsection{Differential Rotation}\label{diffrot}

We carried out a measurement of differential rotation on V410~Tau by repeating the image production process described above, while incorporating a solar differential rotation law of the form 
 
\begin{equation}
\Omega(\rm \theta) = \Omega_{\rm eq} - \Delta\Omega \sin^{2}\theta, 
\label{eqdr}
\end{equation}
 
\noindent where $\Omega(\theta)$ is the angular velocity as a function of latitude ($\theta$),  $\Omega_{\rm eq}$ is the angular velocity at the equator and $\Delta\Omega$ is the difference between the angular velocities at the equator and the poles. 

To constrain the values of $\Omega_{\rm eq}$ and $\Delta\Omega$, we varied their values while holding the information content of the images constant, and produced a 2D plot of $\chi^{2}_{r}$.  For the Stokes I profiles the percentage spot coverage was fixed and for the Stokes V profiles the the magnetic field strength ($\rm B_{aim}$) was held constant. When the I profiles were used we were unable recover the differential rotation parameters. The difficulty here arises because the spot distribution is essentially a polar spot and a band of lower latitude spots between $0-30^{\circ}$, minimising the differences between the resultant images when differential rotation parameters are introduced. 

\begin{figure}
\includegraphics[width=0.35\textwidth,angle=270]{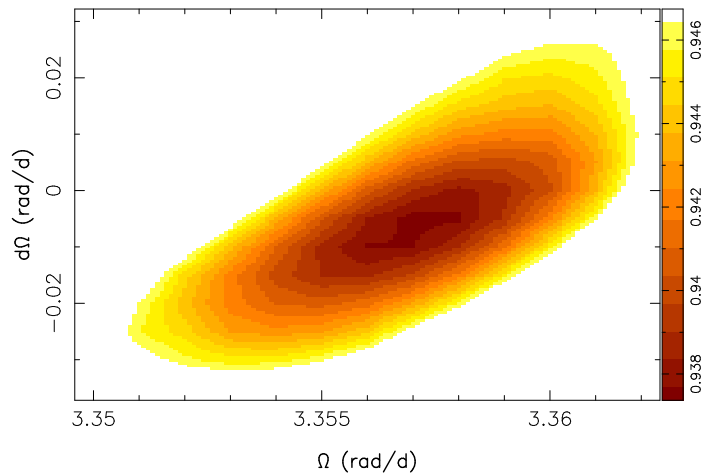}
\caption{Reduced \chisq\ surface used to deduce values of the period and differential rotation. }
\label{diffmap}
\end{figure}

However when the V profiles alone were used in the minimisation it was possible to fit a paraboloid to the \chisq\ plot. The minimum of this paraboloid identifies the $\Omega_{\rm eq}$ and $\Delta\Omega$ favoured by the data. When the code was allowed to converge with a fixed field of 400~G we obtained the \chisq\ surface shown in Fig.~\ref{diffmap}. The minimum of the paraboloid fitted to this surface lies at $\rm \Omega_{\rm eq} = 3.357 \pm 0.001~rad~day^{-1}$ and $\rm\Delta \Omega = -0.007 \pm 0.006~rad~day^{-1}$. This is compatible with solid body rotation as photospheric shear is less than 30\% the solar value at the 3-sigma level. When different values of $\rm B_{aim}$ are used we recover very similar values of $\Omega_{\rm eq}$, $\Delta\Omega$, and their uncertainties.

\subsection{Balmer Line Analysis}\label{ha}

\begin{figure}
\includegraphics[width=0.5\textwidth, clip=true,trim=20 80 0 0]{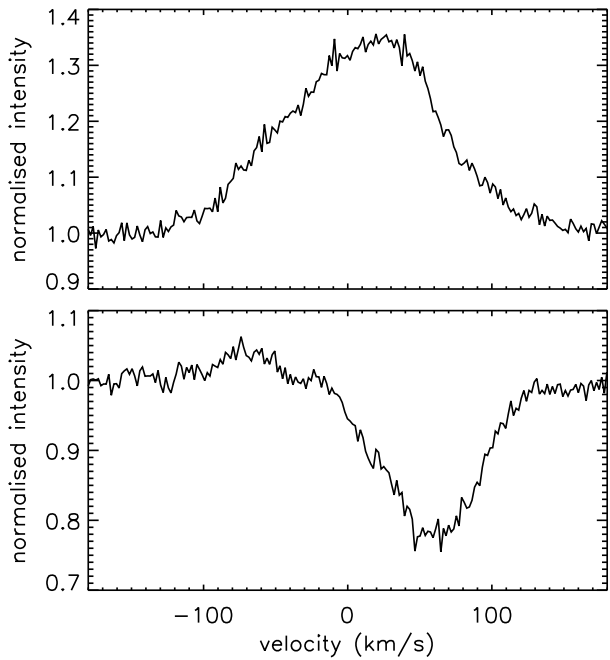}
\caption{Example \ha\ profiles, showing the line in emission at phases 1.522 (top) and in absorption at phase 0.998 (bottom).}
\label{haprofs}
\end{figure}

\begin{figure}
\includegraphics[width=0.5\textwidth]{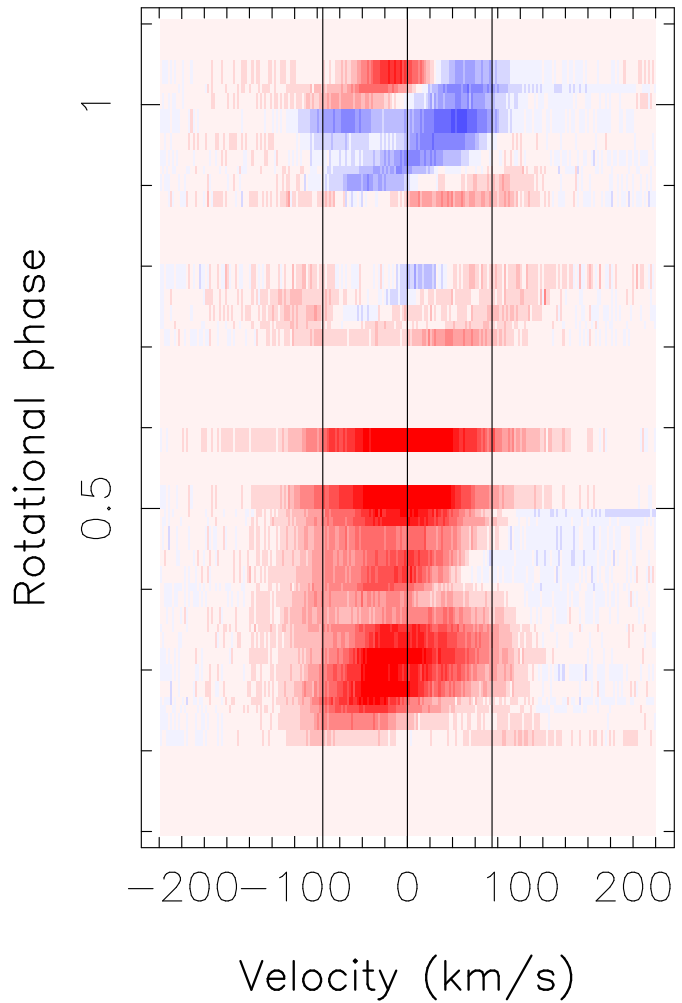}
\caption{Dynamic spectrum of the \ha\ profiles.  Red represents emission and blue is absorption and the normalised intensity range is between 0.7 and 1.3. The vertical lines mark the position of the \vsini\ (75\kms). The phase range on the vertical axis is between 0.1--1.1 to better display the continuation of the absorption feature around phase 1.0.}
\label{dynh}
\end{figure}

The Balmer line emission provides another probe of the surface activity of TTS. The \ha\ profile in wTTS is considered to be a superposition of a photospheric absorption component and a emission component originating in the chromosphere and corona (cTTS have additional contributions from the disc and jet), see e.g \cite{kurosawa06, barrado03}. The \ha\ profiles in V410~Tau, taken from the spectra listed in Tab.~\ref{obs1}, show considerable variability, with the line appearing in both absorption and emission. Example profiles showing this variation are given in Fig.~\ref{haprofs}, one showing the line appearing in emission, the other when it is in absorption. In order to highlight the phase-dependence of the variability Fig.~\ref{dynh} shows the dynamic spectrum. This is created by stacking the profiles in order of phase and displaying as a colour plot. 

The dynamic spectrum shows emission and absorption features. \ha\ is most strongly in emission in the phase range 0.1--0.5, suggesting the presence of chromospheric active regions \citep{v410b} on that hemisphere. These active regions may be connected to the large region of azimuthal field on that hemisphere (see Fig.~\ref{magim}).

This emission feature is crossed by an absorption transient that is centred around phase 0.4. There is a similar absorption transient between phases 0.9--1.05. When such absorption transients have been observed in other stars they have been identified with prominences crossing the stellar disc \citep{cc89, duns06}. The absorption transients appear to be passing through the spectral line more quickly than the spots (i.e. their slopes in Fig.~\ref{dynh} are low), suggesting that they are not very close to the surface. Calculating the variance profile \citep{johns95} reveals a symmetrical profile (not shown), with a full-width half-maximum (FWHM) of 130\kms, and shows variance out to $\pm100$\kms. 

The small amount of variability at velocities larger than \vsini\ appears to suggest that a large fraction of the variability is due to chromospheric activity. While this has been observed previously \citep{fern04}, we currently favour higher-lying features due to the flat slopes of the absorption transients in Fig.~\ref{dynh}.  The absorption feature between phases 0.90 and 1.05 was observed throughout the observation run, i.e. for 8 stellar rotations, suggesting the presence of a stable feature in corotation with the star, such as a prominence. If this is the case we can estimate the height of the prominences from the slopes of the features in Fig.~\ref{dynh}. The absorption transients cross the stellar disc in roughly 0.15 of a rotation. Assuming the prominences are in co-rotation with the star this corresponds to a distance of $2.5~\rm R_{*}\cos\theta$ from the rotation axis, where $\theta$ is the stellar latitude. The co-rotation radius is $2.4~\rm R_{*}$, suggesting they may lie above co-rotation. The equations in \cite{jardine05}  allow stable prominences up to a height of $7.4~\rm R_{*}$.

\section{Discussion}

\subsection{Optical Variability of V410~Tau}\label{pho}

At the present time V410~Tau is displaying a low amplitude optical lightcurve, which we can attempt to relate to the image in Fig.~\ref{dopim}. Previous images \citep{v41094b, hatzes95b, v41096, v410a} showed decentred polar spots and low latitude spots confined to particular longitude ranges. In Fig.~\ref{dopim} the centre of the high latitude spot is very close to the pole and the lower latitude spots are found at all longitudes, leading to less variability as the star rotates. 

\begin{figure}
\includegraphics[width=0.35\textwidth,angle=270]{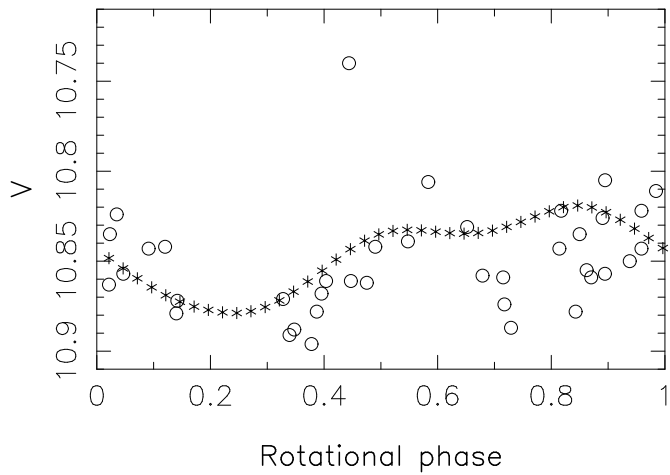}
\caption{V magnitude as a function of rotational phase. The circles show data taken at the Crimean Observatory. The asterisks show the lightcurve calculated by integrating the flux at each phase from the image shown in Fig. \ref{dopim}. The latter has been shifted to have the same mean V magnitude as the photometric measurements. }
\label{photo}
\end{figure}

This variability can be quantified more precisely by comparing with the photometric data and calculating the flux differences that result from the spot distribution in Fig.~\ref{dopim}. The photometric measurements in the V--band made at the Crimean Observatory and described in Sec.~\ref{obs} are shown in Fig.~\ref{photo}. A synthetic lightcurve has been overplotted. This was produced by integrating the flux over the visible surface of the star at each phase. The spot--filling factors were converted to temperatures, using $\rm T_{phot} - T_{spot} = 1400~K$ \citep{petrov94}, and then into fluxes, which are then averaged over the visible surface of the star at each phase (taking limb-darkening into account) and finally converted to a $\rm \Delta V$. This gives a smoothly varying function with a peak brightness at phase 0.85 and a minimum at phase 0.25. The peak-to-peak amplitude of this curve is 0.038 which is roughly consistent with the photometric measurements. 

We believe that the outlying point at phase is associated with a flare. The contemporaneous U--B and B--V colours provide additional evidence for this event. Discounting this point, the dispersion in the measured data points at this time is about 0.02. The mean V magnitude measured during this epoch (10.85) is similar to that measured in \cite{grankin08} where \textless V\textgreater\ lies between 10.8 and 10.95, further supporting the assertion that the change in the lightcurve is associated with a rearrangement in the spots. 

\subsection{Magnetic Nature of V410~Tau}\label{dyn}

Zeeman Doppler images have previously been created for four other TTS, all of which are cTTS. These are BP~Tau \citep{bptau08}, V2129~Oph \citep{v2129}, CR~Cha and CV~Cha \citep{hussain09}.  There is a trend observed in M dwarfs, that fully convective stars have large proportions of their fields concentrated in the lowest $l$ and $m$ modes and stars with radiative cores have more power concentrated in the higher $l$ and $m$ modes \citep{morin07, donati08}. BP~Tau is a fully convective cTTS with a field which is strongly axisymmetric and dipolar, similar to fully convective main sequence stars. V2129~Oph, which has a radiative core, is dominated by an octupolar component which is approximately three times larger than the dipole. CR~Cha and CV~Cha are also partially convective and have significant power in higher $l$ and $m$ modes. The field of V410~Tau has a strong dipolar component, in common with fully convective stars, despite the fact that it appears to be partially convective (see Sec.~\ref{phys}). The large fraction of the field confined into the dipolar modes is significant, given that the high \vsini\ of V410~Tau means that we are sensitive to the higher orders in the spherical harmonic decomposition. 
\begin{figure}
\includegraphics[width=0.35\textwidth, angle=270]{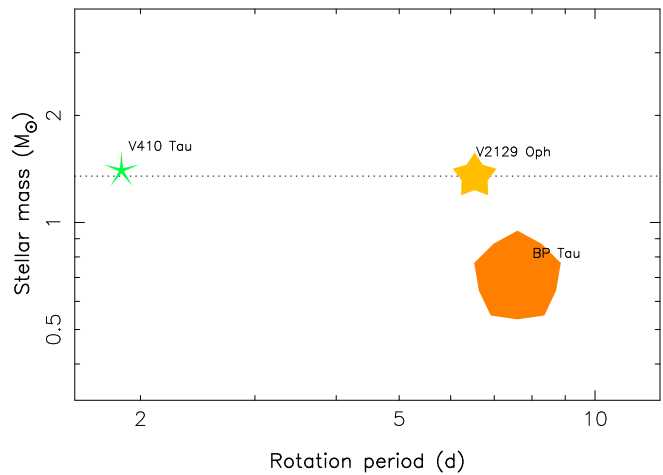}
\caption{Plot of the mass of TTS against rotational period. The size, colour and shape of symbols represent the field strength,  configuration and axisymmetry of the fields respectively. Larger symbols indicate a larger field, red stands for 100\% poloidal and blue stands for 100\% poloidal. A sharp five-pointed star represents a completely non-axisymmetric poloidal field and a decagon represents a completely axisymmetric poloidal field. The dotted line shows the approximate position of the fully convective limit for a 1 Myr star using the \protect\cite{siess00} models. }
\label{bplot}
\end{figure}

Fig.~\ref{bplot} compares the magnetic topology of V410~Tau with BP~Tau and V2129~Oph, using the conventions in \cite{donati08} and \cite{morin08}, who investigated the magnetic topology of M dwarfs. 
While we cannot infer trends from this plot from only three data points it allows direct comparison between these TTS. V2129~Oph and V410~Tau have similar physical structures, with closely matched masses and similar radiative core sizes ($\sim 0.2~\rm R_{*}$ for V2129~Oph).  However, there are significant differences between the fields of the two stars. The field of V410~Tau is more strongly poloidal, and is much less axisymmetric. It is rotating more rapidly than the cTTS BP~Tau and V2129~Oph (due to the absence of disk-braking), but this does not lead to commensurately larger fields as it is the fully convective BP~Tau which generates the largest magnetic field, despite its lower rotation rate. It appears that the depth of the convection zone plays a role in determining the field strength, although we do appear to observe the sharp transition in field topology that is seen in M~dwarfs. We anticipate adding more stars as the data on TTS become available. 

Qualitatively, the field of V410~Tau most resembles older, rapidly rotating stars such as AB~Dor \citep{jf97b, abdor99,donati03, hussain07}, rather than the classical TTS discussed above.
AB~Dor is a 50~Myr old (i.e. post-T~Tauri) K0 star with a \vsini\  of 90\kms. The similarities between the magnetic maps of the two stars are monolithic regions of azimuthal field, almost forming rings around the pole and complex radial fields with many switches in polarity. 

AB~Dor and V410~Tau are broadly similar in the sense that they are young and rapidly rotating. However V410~Tau has almost solid body rotation while AB~Dor has significant differential rotation, with $\Delta\Omega$ (see Eq. \ref{eqdr}) having values between $0.043-0.074 \rm~rad~day^{-1}$ \citep{cc02b, jf03, donati03b, jeffers07}. 

Usually, strong azimuthal field is associated with differential rotation. For instance an $\alpha\Omega$ dynamo gives rise to toroidal field through differential rotation shearing the poloidal field. The field topology of V410~Tau is therefore somewhat difficult to understand. 

While the qualitative similarity of the field with those of partially--convective stars such as AB~Dor suggests that V410~Tau has a radiative core, we cannot rule out the possibility that V410~Tau is fully convective, as Sec.~\ref{phys} suggested. In this case the dipolar component of the field follows the trend observed in other TTS and M~dwarfs. However, the question of the origin of the azimuthal component remains, due to the lack of observed differential rotation. 

Further magnetic and brightness maps will follow from this campaign on V410~Tau. We also have spectra of several other TTS which will be used to produce magnetic images. Using these data, we expect to gain a better insight into the form and evolution of dynamos in pre-main sequence stars. 

\section{Conclusions}

The Doppler image of the weak-line T~Tauri star V410~Tau in January 2009 shows a large polar spot and a several lower latitude spots spread around the latitude band between $0 - 30^{\circ}$. This spot distribution can explain the relatively flat V--band lightcurve currently being observed on the star as, unlike during previous Doppler imaging campaigns, the spots are not confined to a particular longitude range. The polar spot on the image is centred on the pole, unlike on previous images where it was shifted from the pole. 

The \ha\ profile is complex and highly variable. Fast-moving absorption transients in the line may indicate the presence of high-lying circumstellar material, such as slingshot prominences. 

The magnetic field is predominately dipolar and non-axisymmetric. It has a complex radial field as well as a strong azimuthal field with large monolithic regions. The field resembles zero-age main sequence stars such as AB~Dor, but unlike that star it has very weak differential rotation. 
The rotation rate and depth of convection zone appear to be the most important parameters in determining the strength and configuration of the magnetic field

Future Zeeman Doppler images planned as part of a campaign on T\'{e}lescope Bernard Lyot (TBL) will reveal whether changes in the amplitude of the optical lightcurve are accompanied by significant changes in the spot distribution or magnetic field topology.

\section*{Acknowledgments}
 
 We thank the TBL staff for their help during data collection and the anonymous referee for constructive comments that have improved the clarity of the text. This work was supported by the French ``Agence Nationale pour la 
 Recherche'' (ANR) within the ``Magnetic Protostars and Planets'' (MaPP) project.

\bibliographystyle{mn2e}
\bibliography{refs}

\label{lastpage}

\end{document}